\documentstyle[preprint,aps]{revtex}
\begin{document}

\draft
\title{On the gap structure of
UPt$_3$: phases A and B}

\author{
Louis Taillefer$^a$,
Brett Ellman$^a$,
Benoit Lussier$^a$,
Mario Poirier$^b$
}

\address{$^a$Department of Physics,
McGill University,
Montr\'eal, Qu\'ebec, Canada H3A 2T8\\
$^b$D\'epartement de Physique, Universit\'e de Sherbrooke,
Sherbrooke, Qu\'ebec, Canada J1K 2R1}

\date{\today}

\maketitle

\begin{abstract}

We have used
thermal conduction and
transverse sound attenuation
to probe the
anisotropy of the gap structure in two superconducting phases of UPt$_3$.
For the low-temperature phase B, transverse sound has in the past
provided strong evidence
for a line node in the basal plane. Now, from the anisotropy of the thermal
conductivity we further establish the presence of a node
along the c-axis and provide information on its k-dependence. For the
largely unexplored high-temperature phase A, our study of the attenuation for
two directions of the polarization yields
directional information on the quasiparticle spectrum, and the first
clear indication of a different gap structure in the two phases.

\end{abstract}

\pacs{PACS numbers: 74.70.Tx, 74.25.Fy, 74.25.Ld}

\noindent
{\large \bf 1. Introduction}

\vskip 0.5cm

The past decade has seen a tremendous interest in novel superconductors:
low dimensional organic compounds,
high T$_c$
cuprates and
heavy fermion metals \cite{reviews}.
In all of these, the superconducting order parameter is thought to
be unconventional, that is not to have the standard s-wave symmetry.
A typical consequence of non-s-wave symmetry is a gap structure which goes
to zero at certain points on the Fermi surface.
A precise determination
of the position of such nodes, the topology of the gap in their vicinity
and its actual symmetry continue to be one of the central
pursuits in the field.

In this paper, we focus on the heavy-fermion compound UPt$_3$, where
the case for an unconventional superconducting state is compelling: at
ambient pressure and zero magnetic field, the onset of
superconductivity at T$_c^+$=0.5 K is followed by a second transition at
T$_c^-$=0.44 K, as seen in the
specific heat shown in Fig. 1 \cite{Lohneysen}.
The phases above and below T$_c^-$ are called A and B, respectively.
Two related questions arise:
1) how do the order parameters in phases A and B differ?
2) which of the several theoretical scenarios proposed for the superconducting
phase diagram
is correct (if any)
\cite{reviews}?

We have used the propagation of heat and sound to address the first
question, and thereby touch on the second. By virtue of their directional
nature, these two complementary
probes can provide direct
information about the anisotropic gap structure in UPt$_3$.
Details of the experiments can be found in Refs.
\cite{Lussier,Ellman}.

\vskip 0.5cm

\noindent
{\large \bf 2. Results}

\vskip 0.5cm

The crystal structure of UPt$_3$ is hexagonal, and electron transport in the
normal
state is anisotropic with respect to the c-axis.
The electrical resistivity and thermal conductivity
below 0.8 K obey respectively
$\rho$(T)=$\rho_0$+AT$^2$ and
$\kappa_N$(T)=T[a+bT$^2$]$^{-1}$, with
$\rho_0$=0.23 (0.61) $\mu\Omega$ cm,
A=0.59 (1.60) $\mu\Omega$ cm K$^{-2}$,
a=0.09 (0.25) m K$^2$ W$^{-1}$ and
b=0.37 (1.0) m W$^{-1}$,
for a current parallel (perpendicular) to the c-axis \cite{Lussier}.
This implies that:
1) elastic defect scattering in our crystals is low,
2) the Wiedemann-Franz law is verified ($\rho_0$/a=L$_0$=($\pi$k$_B$/e)$^2$/3),
3) inelastic
(electron-electron)
scattering
exceeds elastic scattering just above T$_c$,
4) the anisotropy of $\rho$ and $\kappa$ is independent of temperature, being
the same for elastic and inelastic scattering,
5) the anisotropy is the same for transport of heat and charge,
with conduction along the
c-axis being 2.7 times better.
The normal state is seen to have the characteristics of
a Fermi liquid with an anisotropy of 2.7
in the average Fermi velocity (or mass).

The thermal conductivity is shown in Fig. 1.
With decreasing temperature, the normal state $\kappa$/T
increases and the anisotropy is constant.
At T$_c^+$, and with
the subsequent growth of the condensate,
the number of quasiparticles decreases
and $\kappa$/T goes down dramatically. Therein lies some fruitful information
about the
structure of the gap in the excitation spectrum which we discuss in the next
section.
Note that the development of an additional anisotropy
in the superconducting state is unambiguous
evidence for an anisotropic gap.

The first direct evidence of gap anisotropy in UPt$_3$ came from measurements
of the attenuation $\alpha_{q\epsilon}$(T)
of transverse sound propagating in the basal plane ({\bf q}//a,b)
where two different temperature dependences were observed for a polarization
$\mathbf{\epsilon}$ parallel
and perpendicular to the c-axis \cite{Shivaram 86}.
In these early measurements, a crystal with a single broad transition was used,
so that no information
was obtained about the A-phase. Recently, we repeated the measurement
on a crystal
with two well-defined transitions at T$_c^-$=435 mK and T$_c^+$=495 mK
\cite{Ellman}.
The results are shown in Fig. 1.
For phase B, they agree well with the early data.
The behaviour characteristic of phase A is now revealed,
thanks to the high sensitivity of transverse
sound attenuation to the opening of the gap.
The main finding is the clear difference
in the response of the two phases, which represents
the first direct evidence for a change of superconducting order parameter
at T$_c^-$.

\vskip 0.5cm

\noindent
{\large \bf 3. Discussion}

\vskip 0.5cm

A knowledge of the gap structure
can lead to the group representation of the order
parameter which, in UPt$_3$,
would allow one to discriminate
between the various phenomenological models for the phase diagram.
Depending on the model, the order parameter will belong to either a
one-dimensional (A$_1$, A$_2$, B$_1$, B$_2$) or a two-dimensional
representation
(E$_1$, E$_2$) of the hexagonal group, and have either even (g) or odd (u)
parity
\cite{reviews}.

While there is no consensus at present
on the correct theory of transport in unconventional
superconductors, a generalization of the standard (s-wave)
theory to account for unconventional gap structures
which treats impurity scattering in the unitary limit
is being actively applied to high T$_c$ cuprates and heavy fermion
compounds (see Refs. \cite{Norman 96,Fledderjohann 95,Graf 96}, and references
therein).
Thermal properties such as $\kappa$(T) and $\alpha$(T) calculated within
any theory
will depend on the complex
topology of the Fermi surface and the microscopic pairing interaction.
However, as argued by Graf {\it et al.} \cite{Graf 96},
there is a `diagnostic regime' at low temperature where only a
knowledge of the asymptotic topology of the gap at the nodes is needed.
For that regime,
it should be reasonable
to approximate the Fermi surface by an
ellipsoid
(however, see Ref. \cite{Norman 96}).

Within a simplified picture
of a single ellipsoidal
Fermi surface
with a mass ratio of 2.7 and perfect uniaxial symmetry about the
c-axis, the magnitude of the gap can only depend on the polar angle
$\theta$. A general gap will be a linear combination of ellipsoidal
harmonics Y$_{LM}$ each of which vanishes for one or more values of $\theta$
(except Y$_{00}$).
The nodes can therefore be points at the poles ($\theta$=0), a line
around the equator ($\theta$=90$^\circ$), two lines above
and below
($\theta$=90$^\circ$$\pm$$\theta_0$)
the
equator,
or a combination of these basic elements (see Ref. \cite{Norman 96}).
The five lowest harmonics have the following structure:
Y$_{00}\sim $constant (`s-wave'),
Y$_{10}\sim $cos$\theta$ (`polar'),
Y$_{11}\sim $sin$\theta$ (`axial'),
Y$_{20}\sim $(cos$^2\theta$ - 0.15) (`tropical', since $\theta_0$=23$^\circ$),
Y$_{21}\sim $sin$\theta$cos$\theta$ (`hybrid').
The asymptotic behaviour of the axial gap near the poles, for example, is
linear
(sin$\theta$ $\sim$ $\theta$ for $\theta$ $<$ 20$^\circ$ or so)
and therefore the diagnostic regime corresponds approximately to
k$_B$T$<$$\Delta$($\theta$=20$^\circ$), which translates roughly as
T/T$_c$$<$0.3.

At the lowest temperatures, the theory universally predicts the existence of
impurity-induced low-energy quasiparticle
excitations giving rise to  a `gapless regime', which
corresponds approximately to
k$_B$T$<$2$\hbar\Gamma_0$ \cite{Norman 96,Fledderjohann 95,Graf 96}, where
$\Gamma_0$ is the impurity scattering rate.
In our crystals,
it appears that
$\Gamma_0$=0.05(k$_B$/$\hbar$)T$_c^-$
or less \cite{Lussier}, so that
T/T$_c^-$$<$0.1.

At present, the theory is inadequate in treating
electron-electron scattering, and a meaningful comparison with experiment
should be
limited to the `elastic regime'.
For our UPt$_3$ crystals, this implies
T/T$_c^-$$<$0.3-0.4.

\vskip 0.5cm

{\large \bf Phase B}

\vskip 0.5cm

By concentrating on the interval 0 $<$ T $<$ 0.3 T$_c^-$,
we can hope for a powerful diagnostic
on the nodal structure of phase B,
free of the complicating effects of electron interactions and
gaplessness.

The rise in $\kappa$/T and $\alpha$ from T=0 for a heat current and a sound
polarization
perpendicular to the c-axis is much more rapid than in conventional
superconductors.
In particular, the linear behaviour of
$\alpha$(T)
roughly down to T/T$_c^-$=0.1 \cite{Shivaram 86}
is strong evidence
for a gap vanishing in (or near)
the basal plane. For a uniaxial gap structure, this means a line
node around (or near) the equator.
Note that because transverse sound is mostly attenuated by quasiparticles with
wavevectors
neither perpendicular to {\bf q} nor to
$\mathbf{\epsilon}$ \cite{Ellman}, $\alpha_{ac}$
does not pick out this line node very much.
Moreover, neither $\alpha_{ac}$ nor $\alpha_{ab}$ are expected to be very
sensitive to possible nodes along the c-axis.
As a result, transverse sound in the B-phase of UPt$_3$ has been
qualitatively interpreted as evidence for a gap with a line node in the basal
plane
\cite{reviews}.
This eliminates 2 of the 5 gaps listed above. Indeed,
neither the s-wave gap nor the axial gap vanish in (or near)
the basal
plane. Whether the data can allow one to distinguish between the polar gap
and a hybrid gap, for example, requires a model calculation.
It is clear, though, that transverse sound
is not ideal for probing the gap near $\theta$=0.

In this respect, the conduction of heat is better suited,
being dominated by quasiparticles travelling in the forward direction.
For example, the presence of a node at $\theta$=0 will enhance the
quasiparticle current along the c-axis in a hybrid gap relative to the polar
gap.
This possibility is best investigated with the anisotropy ratio
$\kappa_c$/$\kappa_b$,
plotted in Fig. 2.
The striking result is that
$\kappa_c$/$\kappa_b$
extrapolates to a finite value at T=0, about half that of
the normal state.
By inspection, one can deduce the limiting value of
$\kappa_c$/$\kappa_b$
as T$\rightarrow$0 for 3 of the 5 gaps above:
the s-wave with
$\Delta_c$$>$$\Delta_b$
and the polar gap must go to zero,
the s-wave with
$\Delta_c$$<$$\Delta_b$
and any axial gap to
infinity \cite{Lussier}.
Hence, the anisotropy of heat conduction not only confirms that the gap
of phase B is most definitely not s-wave or axial,
but it also shows
quite straightforwardly that it is not polar.
Of particular interest are the two lowest hybrid gaps
($\sim$sin$^n\theta$cos$\theta$,
n=1,2),
because they
correspond to two of the states most often
postulated for phase B \cite{reviews}.
They belong
respectively to the E$_{1g}$ and E$_{2u}$ representation and their
overall structures are very similar except near the poles, where the gap
opens up linearly in E$_{1g}$ and quadratically in E$_{2u}$ \cite{Norman
96,Fledderjohann 95}.
Since the density of states
N(E)$\sim$E
for a line node or a quadratic point node,
while N(E)$\sim$E$^2$ for a linear point node, it is natural to expect
$\kappa_c$/$\kappa_b$
to remain finite as T$\rightarrow$0 in E$_{2u}$ and go to zero in E$_{1g}$,
as first shown by Fledderjohann and Hirschfeld \cite{Fledderjohann 95}.

In Fig. 2 we compare the data with calculations using resonant impurity
scattering
with $\Gamma_0$=0.05T$_c^-$, for the two hybrid gaps. The basic gaps
Y$_{21}$ and
Y$_{32}$
(the lowest
harmonics allowed by symmetry)
are mixed with a small amount of
the next harmonics of the same symmetry to optimize the fit \cite{Norman 96}.
$\kappa_b$/T is well reproduced by the calculation for both gaps.
It is along the c-axis that the gaps
differ and the disparity in the behaviour
of the two gaps is dramatically brought out by looking at the ratio of
$\kappa_c$ and $\kappa_b$. The data
for $\kappa_c$/$\kappa_b$
is almost flat and extrapolates
to a value of 0.4 to 0.5 at T=0, as also found by Huxley {\it et al.}
\cite{Huxley 95},
something which the E$_{2u}$ gap can easily reproduce.
On the other hand, the E$_{1g}$ gap above the gapless regime is
qualitatively different, being characterized by
a smooth
extrapolation to zero. If the gapless regime is suppressed by reducing
$\Gamma_0$,
the calculated
ratio does eventually go to zero \cite{Norman 96,Fledderjohann 95}, as expected
on simple
grounds of topology.
We conclude that the anisotropy of heat conduction favours a hybrid gap of
E$_{2u}$
symmetry over one of E$_{1g}$ symmetry for phase B of UPt$_3$,
at least within our ellipsoidal model.

\vskip 0.5cm

{\large \bf Phase A}

\vskip 0.5cm

Let us now turn to phase A, for which very little is known
as a result of its limited range in temperature.
The most significant information comes from
the transverse ultrasound attenuation data \cite{Ellman}, shown in Fig. 1.
As may be seen, $\alpha_{ab}$ drops initially with
decreasing temperature before
becoming roughly constant, while $\alpha_{ac}$ only has a slight "bump"
seemingly superimposed on
the sharply falling attenuation observed in the B-phase.
Qualitatively, this implies
that more quasiparticles exist in phase A than would be present if
phase B extended up to the same temperature.  Furthermore, it appears that
these extra excitations preferentially scatter sound when the polarization is
in the basal plane.
To see this we plot, in Fig. 3, the data of Fig. 1 normalized to the
attenuation
at either $T_c^+$ or $T_c^-$
as a function of temperature normalized to the appropriate
critical temperature.  This allows us to compare, say, the B-phase
attenuation (for either polarization) with the attenuation in the A-phase
over the same reduced temperature range.  It is evident that $\alpha_{ab}$
is much enhanced in the A-phase as compared to the B-phase.  The data for the
c-axis polarization, $\alpha_{ac}$,
however, are roughly equal in the two phases.

The data
contain precise information on the
momentum space distributions of quasiparticles, and thus on the gap structure.
Because we are not in the `diagnostic regime',
such information can only be extracted by comparison with complete
calculations,
analogous to those for the thermal conductivity.
It must be stressed, however, that quantitative interpretations of the A-phase
data assuming a specific gap must carefully consider the effect
of quasiparticle-quasiparticle interactions,
since the structure and evolution of the gap at high temperatures affect the
quasiparticle density and therefore the magnitude of the inelastic scattering.
We emphasize, however, that while changes in inelastic scattering due to
superconductivity might affect the {\it size} of the observed anisotropy,
this anisotropy still derives from that of the gap and thus reflects the
symmetry of the order parameter.

We therefore conclude from the observed difference in the anisotropy of the two
phases,
that the order parameter associated with phase A must change upon going into
phase B.
This is the first solid evidence for a transition between two distinct
superconducting states at T$_c^-$.

\pagebreak

\noindent
{\large \bf Conclusions}

\vskip 0.5cm

In summary, measurements of the thermal conductivity of UPt$_3$
down to T$_c$/10 have
shed light on
the nodal structure of the gap function in the low-temperature phase B.
The unusual observation of a finite value for the anisotropy ratio
$\kappa_c$/$\kappa_b$ as
T$\rightarrow$0 leads to new information:
the gap vanishes along the c-axis, and it does so with a special
angular dependence compatible with
E$_{2u}$ but not with E$_{1g}$ symmetry, within an ellipsoidal model
for the Fermi surface, as confirmed
by calculations
based on resonant impurity scattering \cite{Norman 96}.

The attenuation of transverse ultrasound in
a crystal of UPt$_3$ with two well-defined transitions
at T$_c^-$ and T$_c^+$ shows the gap structure
of phase A to be qualitatively different to that of phase B,
with a significant directionally-dependent enhanced density of
quasiparticles.  In combination with resonant
impurity scattering calculations which take into account the normal state
properties and the electron-electron interactions,
these data should provide some of the first constraints on the
symmetry of the A-phase.

\vskip 0.5cm

\noindent
{\large \bf Acknowledgements}

\vskip 0.5cm

We are grateful to M.R. Norman and P.J. Hirschfeld
for extensive
discussions and for allowing us to reproduce their calculations,
and to H. von L\"ohneysen and M. Sieck for the specific heat data.
This work
was funded by NSERC of Canada and FCAR of Qu\'ebec. L.T.
acknowledges the support of the Canadian Institute for Advanced Research and
the A.P. Sloan Foundation.

\begin{figure}
\caption{
Thermal properties of UPt$_3$ in phases A and B
of the superconducting state, normalized to their
value in the normal state (at 0.53 K).
Top panel:
specific heat divided by temperature, showing the two distinct transitions
at T$_c^-$=435 mK and T$_c^+$=495 mK (vertical dashed lines) [2].
Middle panel:
thermal conductivity divided by temperature for a heat current both parallel
and
perpendicular to the hexagonal axis, showing the appearence of additional
anisotropy
below T$_c^-$ [3].
Bottom panel:
attenuation of transverse ultrasound propagating in the basal plane for a
polarization both
parallel and
perpendicular to the hexagonal axis, showing
a distinct change in going from phase A to phase B [4].
}
\label{fig1}
\end{figure}

\begin{figure}
\caption{Top panel: low-temperature thermal conductivities
along axes c
($\kappa_c$; open circles) and b ($\kappa_b$; solid circles)
vs reduced
temperature and normalized to 1 at T$_c^-$ [3].
Bottom panel: the anisotropy ratio $\kappa_c$/$\kappa_b$.
The data are compared with calculations [6]
for hybrid gaps
in E$_{1g}$ (dashed lines) and E$_{2u}$ symmetry (solid lines).
The impurity scattering rate $\Gamma_0$ is taken to be 0.05 T$_c^-$.
}
\label{fig2}
\end{figure}

\begin{figure}
\caption{
Transverse attenuation data for both polarizations normalized to the
value of the attenuation at T$_c^+$
and T$_c^-$ as a function of T/T$_c^{+,-}$.  This choice of
normalization allows us to compare the attenuation in the A and B phases
over the same (reduced) temperature range.  The A-phase shows an enhanced
in-plane attenuation vs. the B-phase while the out-of-plane polarization
data are roughly equal in the two phases.  The lines are guides to the
eye (after Ref. [4]).}
\label{Fig3}
\end{figure}

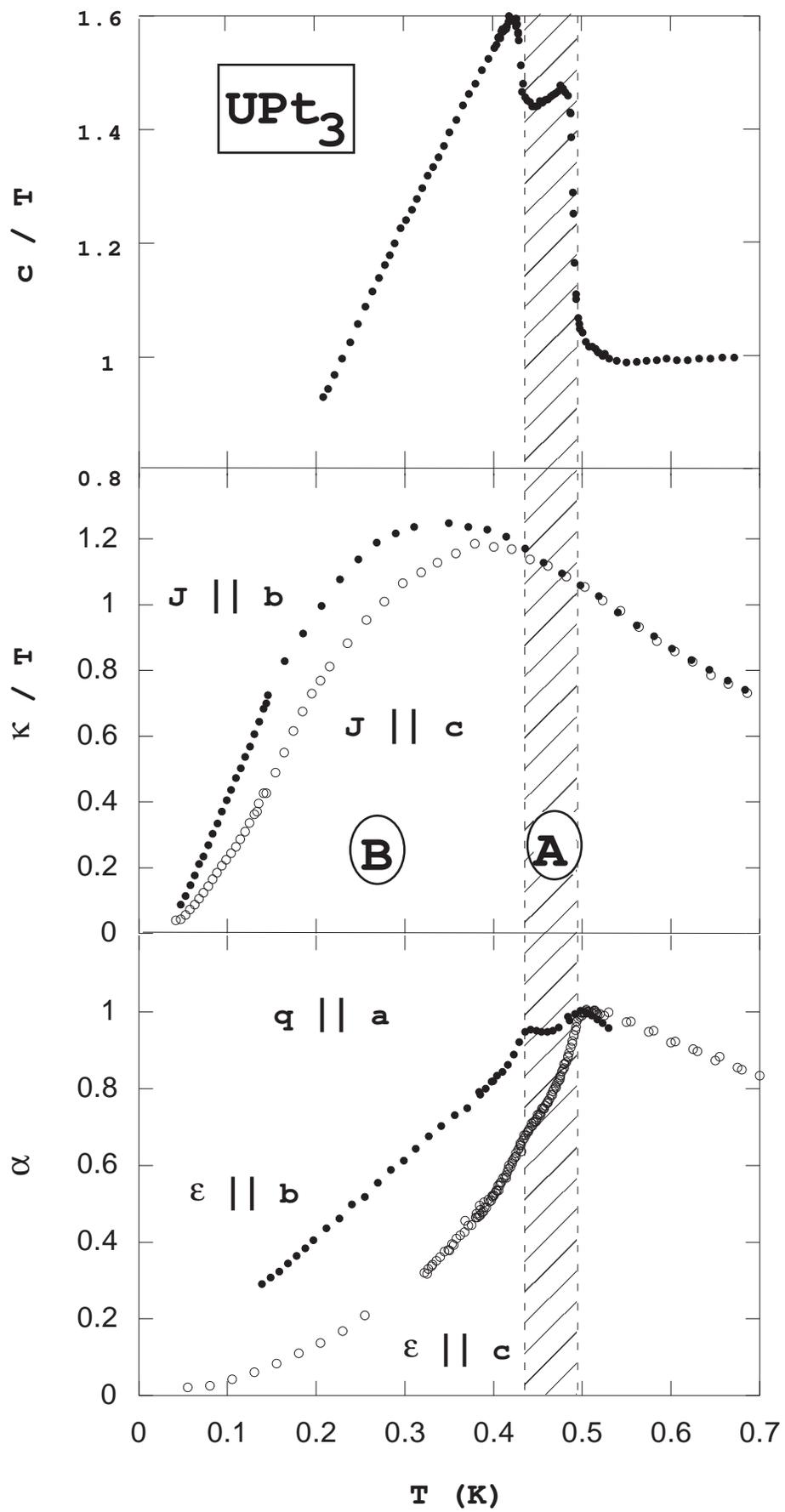
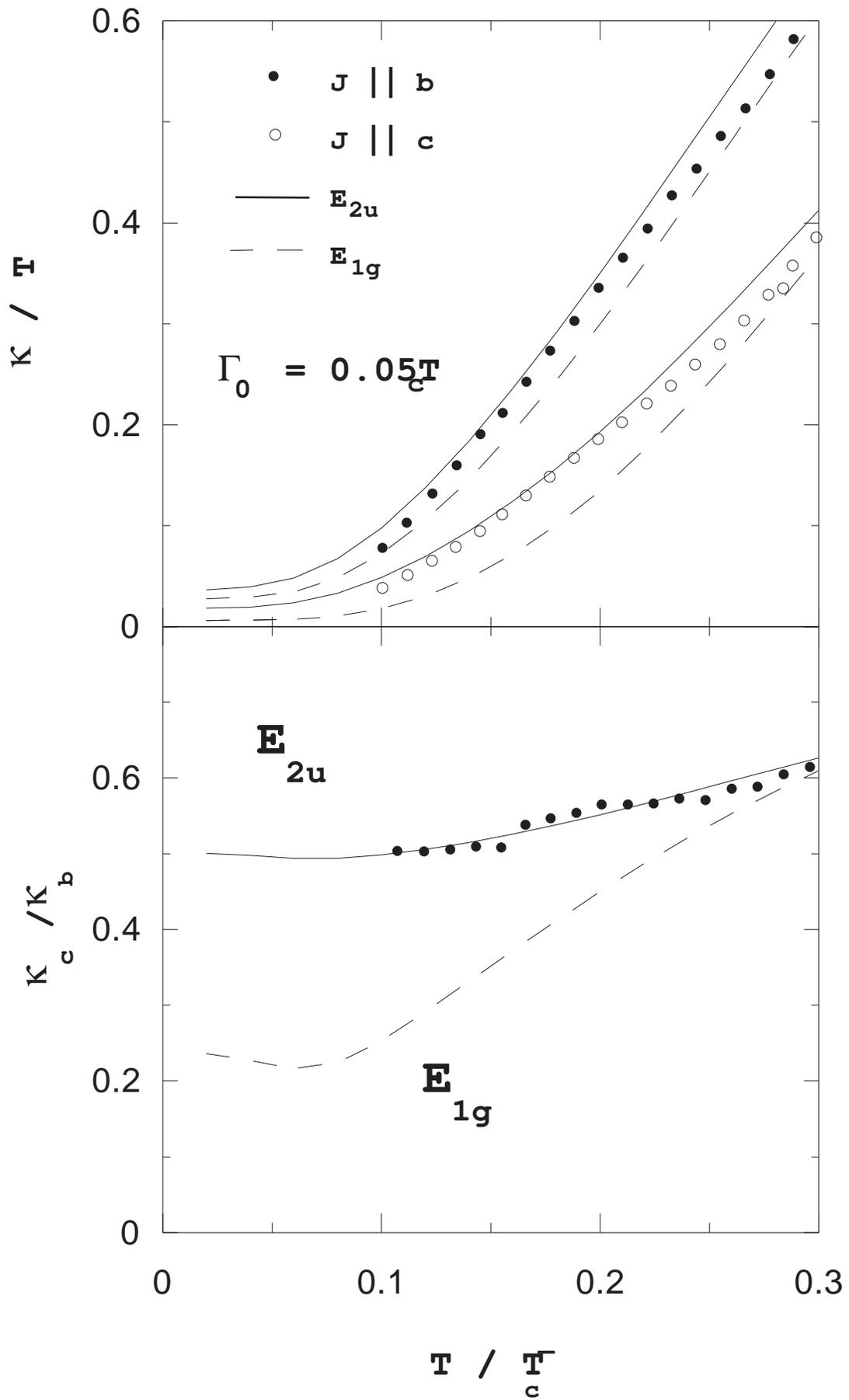
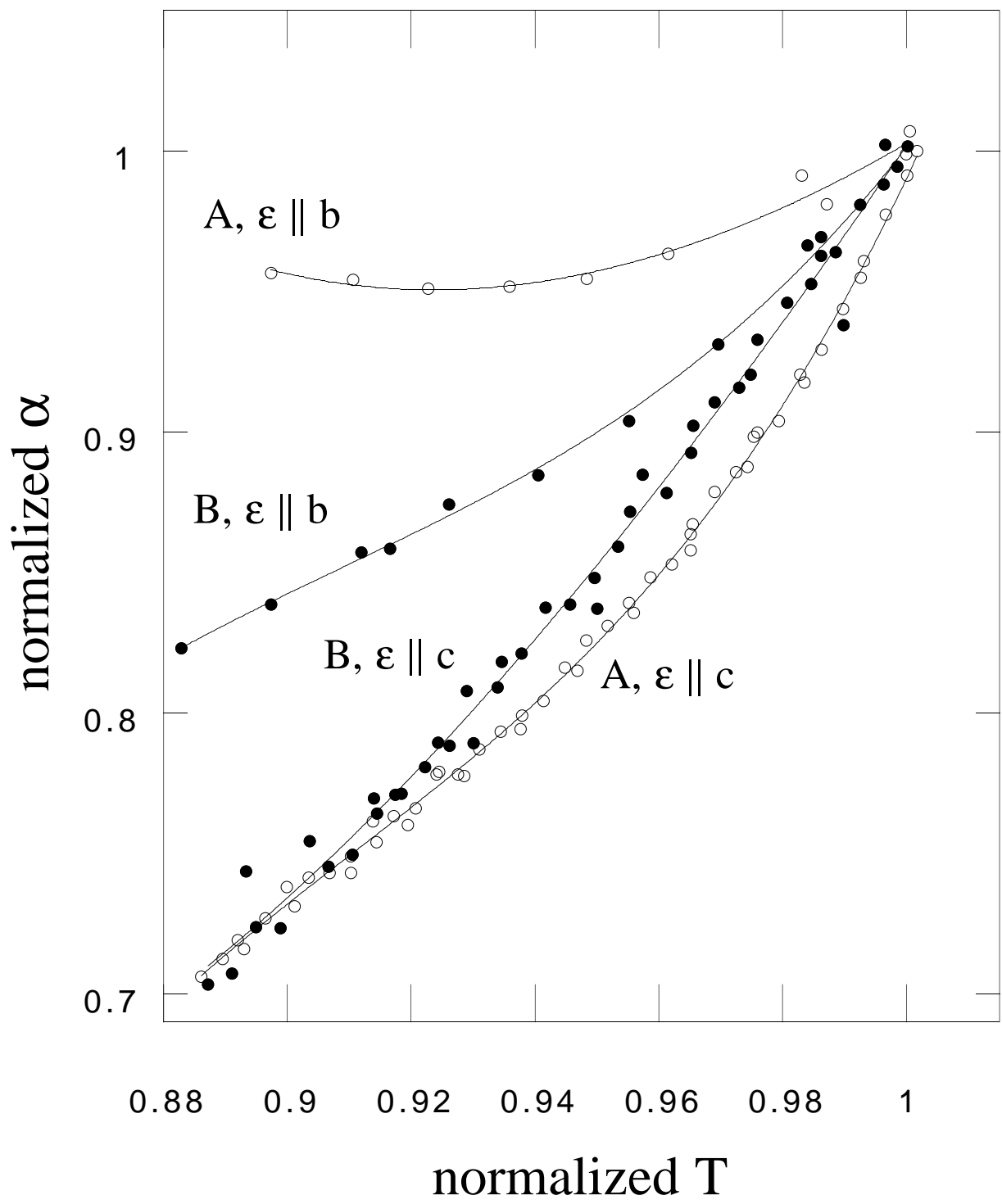
\end{document}